\documentclass{JHEP3} 

\usepackage{epsfig,multicol}
\usepackage{amsmath, amssymb}
\usepackage{epic}
\usepackage{concmath, palatino}

\newcommand\fverb{\setbox\pippobox=\hbox\bgroup\verb}
\newcommand\fverbdo{\egroup\medskip\noindent%
			\fbox{\unhbox\pippobox}\ }
\newcommand\fverbit{\egroup\item[\fbox{\unhbox\pippobox}]}
\newbox\pippobox

\newcommand{\nnn}{{\cal N}=4}
\newcommand{\be}{\begin{equation}}
\newcommand{\ee}{\end{equation}}
\newcommand{\ba}{\begin{eqnarray}}
\newcommand{\ea}{\end{eqnarray}}

\newcommand{\refeq}[1]{Eq.~(\ref{eq:#1})}
\newcommand{\refsec}[1]{Section~\ref{sec:#1}}

\newcommand{\ads}{AdS_5\times S^5}

\title{The virtual scaling function of twist operators \\ in the ${\cal N}=6$ Chern-Simons theory}

\author{Matteo Beccaria and Guido Macorini\\
  Physics Department, Salento University, 
  Via Arnesano, 73100 Lecce\\
  INFN, Sezione di Lecce\\
  E-mail: \email{matteo.beccaria@le.infn.it} 
}

\abstract{
We consider twist-$L$ operators in the planar ${\cal N}=6$ superconformal Chern-Simons ABJM theory. Their anomalous dimension $\gamma_L^{\rm CS}(N)$ 
is a function of the twist $L$, the spin $N$, and the dressed coupling of ABJM. We show that at next-to-leading order in the large spin expansion, 
this anomalous dimension is related to that of $\nnn$ SYM twist operators by a simple scaling law. 
}

\begin{document}

\section{Introduction}

The maximally supersymmetric $\nnn$ SYM theory is an ideal theoretical laboratory for the investigation of planar gauge theory.
At weak coupling, all-order integrability leads to long range Bethe Ansatz equations which are an efficient computational tool
and replace the usual diagrammatical expansion~\cite{Beisert:2005fw}. At strong coupling, AdS/CFT correspondence provides a non trival 
reformulation in terms of type IIB superstring propagating on $\ads$~\cite{Maldacena:1997re}. This leads to quantitative predictions which are 
hardly achievable by different means.

In this perspective, the so-called twist operators are an important probe of various universal quantities characterizing the gauge dynamics. 
At weak coupling, they are gauge invariant single trace composite operators 
belonging to a perturbatively closed $\mathfrak{sl}(2)$ sector. They are built as a matrix product of $L$ fundamental $\nnn$ SYM scalars $\varphi$
with a certain number $N$ of covariant derivatives spread along the chain, ${\cal O}_{L, N} \sim \mbox{Tr}(\varphi^{L-1}\,{\cal D}^N\,\varphi)$. 
The parameters $L$ and $N$ are called the twist and spin of the operator for clear reasons. 
The dual state at strong coupling is well identified in twist-2 and is described classically by a rotating folded string 
configuration~\cite{Gubser:2002tv,Frolov:2002av}. 

The analysis of twist operators is based on the study of their large spin limit, $N\to\infty$. In this regime, we can study their anomalous dimension 
$\gamma_{L,N}(g)$ as a function of the twist $L$ and, of course, the planar coupling $g$. The next-to-leading (NLO) expansion of $\gamma_{L,N}(g)$ 
has the general simple form~\cite{Korchemsky:1992xv,Belitsky:2003ys,Belitsky:2006en,Alday:2007mf}.
\be
\label{eq:NLO}
\gamma_{L, N}(g) = f(g)\,(\log\,N+\gamma_{\rm E}-(L-2)\,\log\,2)+B_L(g) + \cdots,
\ee
where we have neglected terms vanishing at large $N$~\footnote{
The subleading corrections to \refeq{NLO} are very interesting and are related to a generalized Gribov-Lipatov reciprocity
as first discussed in \cite{Dokshitzer:2005bf,Dokshitzer:2006nm,Basso:2006nk}. It is a property very well tested at weak 
coupling~\cite{ReciprocityTests}, and partially investigated in string theory\cite{Beccaria:2008tg}.
}. The coupling dependent coefficient of the logarithm
is equal to twice the conventional QCD cusp anomaly~\cite{Belitsky:2006en} and is commonly referred as the {\em scaling function}.
It is an universal quantity which is known at four loops by diagrammatical methods~\cite{Bern:2006ew,Cachazo:2006az}.
An integral equation for $f(g)$ has been proposed in~\cite{Beisert:2006ez} (BES). It  can generate the
weak coupling expansion of $f(g)$ at any desired order. The strong coupling expansion of $f(g)$ is also known~\cite{Basso:2007wd} (see also \cite{Kostov:2008ax}).

From a physical point of view, the cusp anomaly defines what is called the physical coupling which measures the (universal) intensity of 
soft gluon radiation~\cite{Dokshitzer:2006nm}. On the other hand, the physical meaning of the subleading constant term $B_L(g)$ is more subtle. It 
has been partially clarified in the recent analysis of Dixon-Magnea-Sterman~\cite{Dixon:2008gr} (DMS). From the work of DMS, we know that 
$B_L$ is related to a piece of the infrared divergence of on-shell scattering amplitudes in $\nnn$ SYM. 
The singular part of the amplitudes can be written schematically in the following factorized form~\cite{Infrared}
\be
{\cal A}_{\rm IR} \sim \exp\left(\frac{f^{(-2)}(g)}{\varepsilon^2}-\frac{G^{(-1)}(g)}{\varepsilon}\right),
\ee
where $\varepsilon$ is the IR dimensional regulator and $F^{(-n)}$ is the n-th times iterated logarithmic integral. The new function $G(g)$ is 
the collinear anomalous dimension~\cite{Bern:2005iz,Cachazo:2007ad}. The important remark of DMS is that (at least for twist-2) $G$
is the sum of a universal contribution $G_{\rm eik}$ and the subleading constant $B_2(g)$. The universal part is an eikonal contribution
which can be extracted from the soft logarithms of the Drell-Yan process or from a suitable rectangular light-like Wilson loop.
The non-universal piece $B_2(g)$ comes from the virtual contribution to the Altarelli-Parisi splitting kernel~\cite{Altarelli:1981ax}.
In Mellin space, this is precisely the subleading constant term appearing in~\refeq{NLO}. We shall call it the {\em virtual scaling function}.
The DMS proposal has been recently confirmed at strong coupling in~\cite{Alday:2009zf}.

For the case of $\nnn$ SYM, an integral equation for $B_L(g)$ has been analyzed in~\cite{Freyhult:2009my,Fioravanti:2009xt}.
The equation is derived neglecting wrapping effects. These are well known at leading order for $\mathfrak{sl}(2)$ twist operators~\cite{Bajnok:2008qj}.
At weak coupling, they do not affect the constant $B_L(g)$. Also at strong coupling, this turns out to be true as an outcome of the analysis of~\cite{Freyhult:2009my}.

Recently, twist operators have also been introduced and studied in the ABJM theory~\cite{Aharony:2008ug} which has an integrable structure quite close to that 
of $\nnn$ SYM, despite being a very different theory. ABJM is a 3d $U(N)\times U(N)$ Chern-Simons gauge theory with opposite levels $+k$, $-k$.
It has ${\cal N}=6$ superconformal symmetry and describes the low energy limit of 
$N$ parallel M2-branes at a $\mathbb{C}^4/\mathbb{Z}_k$ singular point. For large $N, k$ and fixed ratio $\lambda=N/k$, we identify $Z_k\simeq S^1$
and M theory on the gravity dual orbifold $AdS_4\times S^7/\mathbb{Z}_k$ reduces to type IIA string on $AdS_4\times \mathbb{CP}^3$. The classical string 
integrability~\cite{Arutyunov:2008if,Stefanski:2008ik,Gomis:2008jt} has a gauge theory quantum counterpart first analyzed in~\cite{Minahan:2008hf,Bak:2008cp}.
In~\cite{Gromov:2008qe}, a set of all-loop Bethe-Ansatz equations have been proposed for the full $\mathfrak{osp}(2,2|6)$ theory
consistent with the string algebraic curve at strong coupling~\cite{Gromov:2008bz}. The equations depend on a dressed coupling $h(\lambda)$ which takes into account the fact that 
the one-magnon dispersion relation is not protected by supersymmetry~\cite{Nishioka:2008gz,Gaiotto:2008cg,Grignani:2008is,McLoughlin:2008he}. The conjectured $S$-matrix
is worked out in~\cite{Ahn:2008aa}.

ABJM twist operators and their anomalous dimensions are discussed in~\cite{Gromov:2008qe,Zwiebel:2009vb,Beccaria:2009ny}. The similitudes with $\nnn$ SYM
are many. At strong coupling, the dual string state is a folded string 
rotating in $AdS_3$ with large spin $N$ and with angular momentum $J\sim \log\,N$ in $\mathbb{CP}^3$~\cite{folded3} not so different than the $\ads$ case.
At weak coupling, their anomalous dimension is captured by all-loop Bethe Ansatz equation which are precisely those of $\nnn$ SYM apart from 
a phase twist. This phase deformation~\cite{Frolov:2005ty} changes drastically some known fine features which are 
shared by $\nnn$ SYM and QCD (Low-Burnett-Kroll wisdom, Gribov-Lipatov reciprocity). 
Nevertheless, the scaling function of ABJM is simply halved, apart from the unavoidable replacement $g\to h(g)$, and we have~\cite{Gromov:2008qe}
(see also \cite{Gromov:2008fy})
\be
f^{\rm CS}(h) = \frac{1}{2}\,f^{\nnn}(h),
\ee
where, here and in the following, CS stands for Chern-Simons and identifies the ABJM case.
This factor $1/2$ can be  understood in terms of the mode number change due to the phase twist in the (thermodynamical) large spin limit.

One immediately asks the following questions. What about the subleading constants appearing in~\refeq{NLO} ? How do they change in ABJM ?  
Is there a simple relation with the $\nnn$ SYM values ? 
The result of this paper is that at next-to-leading order, {\em i.e.} at the level of the expansion \refeq{NLO}, we can write 
\be
\label{eq:result}
\gamma_L^{\rm CS}(N) = \frac{1}{2}\,\gamma^{\nnn}_{2\,L}(2\,N), \qquad (\mbox{NLO}).
\ee
In other words, we predict
\be
\label{eq:final}
\gamma^{\rm CS}_{L, N}(h) = \frac{1}{2}\,f^{\nnn}(h)\,(\log\,(2\,N)+\gamma_{\rm E}-2\,(L-1)\,\log\,2)+\frac{1}{2}\,B^{\nnn}_{2L}(h) + \cdots.
\ee
This conclusion is derived under the same hypothesis about wrapping we assumed in $\nnn$ SYM~\footnote{A first positive test of this assumption
has been presented in~\cite{Beccaria:2009ny} for twist-1 and twist-2 operators at weak coupling and leading wrapping order.}.
Notice that \refeq{result} is false beyond NLO order. 

\medskip
The plan of the paper is as follows. In \refsec{twistops}, we briefly recall a few necessary basic definitions. In \refsec{fourth},
we present a new closed formula for the (asymptotic) anomalous dimension of ABJM twist-2 operators at fourth order, {\em i.e.} eight loops.
In \refsec{experiments}, we compute and collect our results for the NLO large spin expansion of the ABJM anomalous dimensions.
In \refsec{N4}, we briefly recall the form of the NLO BES equation in $\nnn$ SYM as a necessary preliminary steps for the introduction of its
modified version valid in ABJM. \refsec{baxter} illustrates the analysis of the one-loop ABJM Baxter equation which describe twist operators. 
The analysis is aimed at providing arguments for the scaling behaviour of the hole solutions to the Bethe Ansatz equations. 
In \refsec{newintegral}, we report the $\nnn$ and ABJM NLO BES equations in unified form and prove the scaling relation as a simple consequence.
Finally, \refsec{comments} is devoted to some final comments.

\section{Twist operators in ABJM and their exact anomalous dimensions}
\label{sec:twistops}

The all-loop Bethe equations for ABJM has been proposed in~\cite{Gromov:2008qe}. They are associated with the 
$\mathfrak{osp}(2,2|6)$ Dynkin diagram (in the fermionic $\eta=-1$ grading)
\be
\begin{minipage}{260pt}
\setlength{\unitlength}{1pt}
\small\thicklines
\begin{picture}(160,80)(-60,-50)
%
\put(  0,00){\circle{15}}
\put( -5,-5){\line(1, 1){10}}  
\put( -5, 5){\line(1,-1){10}}  
\put(  0,-15){\makebox(0,0)[t]{$u_1$}} 
\put(  7,00){\line(1,0){26}} 
\put( 40,00){\circle{15}}     
\put( 40,-15){\makebox(0,0)[t]{$u_2$}} 
\put( 47,00){\line(1,0){26}} 
\put( 80,00){\circle{15}}
\put( 75,-5){\line(1, 1){10}}  
\put( 75, 5){\line(1,-1){10}}  
\put( 80,-15){\makebox(0,0)[t]{$u_3$}}  
\put( 88,00){\line(3,2){25}} 
\put(120,20){\circle{15}}
\put(115,15){\line(1, 1){10}}  
\put(115,25){\line(1,-1){10}}  
\put( 88,00){\line(3,-2){25}} 
\put(120,-20){\circle{15}}
\put(115,-25){\line(1, 1){10}}  
\put(115,-15){\line(1,-1){10}}  
\put(150,18){\makebox(0,0)[b]{$u_4$}}
\put(150,-28){\makebox(0,0)[b]{$u_{\overline 4}$}}
\put(118,13){\line(0,-1){25}} 
\put(122,13){\line(0,-1){25}} 
\end{picture}
\end{minipage}
\ee
Twist operators in the $\mathfrak{sl}(2)$ sector are obtained by exciting 
the same number $N$ of $u_4$ and $u_{\overline{4}}$ roots. As in the ${\cal N}=4$ case, we shall refer to 
the integer $L$ as the twist of the operator. More details can be found in~\cite{Zwiebel:2009vb}. 

Bethe Ansatz equations are written in terms of the deformed spectral parameters $x^\pm(u)$ defined by 
\be
x^\pm + \frac{1}{x^\pm} = \frac{1}{h}\left(u\pm \frac{i}{2}\right),
\ee
where $h(\lambda)$ is the interpolating coupling entering the one-magnon dispersion relation. 
For twist $L$ operators they are
\be
\left(\frac{x^+_k}{x^-_k}\right)^L = -\prod_{j\neq k}^N\frac{u_k-u_j+i}{u_k-u_j-i}\,\left(
\frac{x_k^--x_j^+}{x_k^+-x_j^-}\right)^2\,\sigma^2_{\rm BES}.
\ee
The only difference compared with ${\cal N}=4$ SYM is the extra minus sign whose effects are discussed in ~\cite{McLoughlin:2008he,Beccaria:2009ny}.
The factor $\sigma_{\rm BES}$ is the Beisert-Eden-Staudacher dressing phase. The momentum constraint is automatically satisfied for 
Bethe root distributions symmetric under $u\to -u$ as those we are interested in.

The contribution to the energy/anomalous dimension from the Asymptotic Bethe Ansatz (ABA) equations is conveniently 
written in terms of  
\be
p(u) = -i\,\log\frac{x^+(u)}{x^-(u)}, \qquad 
u(p) = \frac{1}{2}\,\cot\frac{p}{2}\,\sqrt{1+16\,h^2\,\sin^2\frac{p}{2}},
\ee
and reads
\be
\gamma^{\rm CS}_L(N, h) = \sum_{k=1}^N\left[\sqrt{1+16\,h^2\,\sin^2\frac{p_k}{2}}-1\right] = \sum_{n=1}^\infty \gamma^{\rm CS}_{L, 2\,n}(N)\,h^{2\,n}.
\ee
From the results of~\cite{Beccaria:2009ny}, we know the closed form of the {\em asymptotic} anomalous dimension at three orders for both twist 1 and 2.
We recall here the expressions for the reader's benefit. With the usual definition of  (nested) harmonic sums
\be
S_a(N) = \sum_{n=1}^N \frac{(\mbox{sign} a)^n}{n^{|a|}},\quad S_{a, b, \dots}(N) = \sum_{n=1}^N \frac{(\mbox{sign} a)^n}{n^{|a|}}\,S_{b, \dots}(n).
\ee
we have for twist 1
\ba
\gamma^{\rm CS}_{1, 2}(N) &=& 4\,\left(S_1-S_{-1}\right), \\
\gamma^{\rm CS}_{1, 4}(N) &=& -16 (S_{-3}-S_3+S_{-2,-1}-S_{-2,1}+S_{-1,-2}-S_{-1,2}-S_{1,-2}+S_{1,2}-S_{2,-1}+S_{2,1}+\nonumber\\
&& + S_{-1,-1,-1}-S_{-1,-1,1}-S_{1,-1,-1}+S_{1,-1,1}),
\ea
and a much longer expression for $\gamma^{\rm CS}_{1, 6}(N)$ which can be found in~\cite{Beccaria:2009ny}. All harmonic sums have in this case argument $S_\mathbf{a}\equiv S_\mathbf{a}(N)$.
For twist-2, we find instead the compact expressions
\ba
\gamma^{\rm CS}_{2, 2}(N) &=& 4\,S_1, \\
\gamma^{\rm CS}_{2, 4}(N) &=& 4\, S_3-8\, S_{1,2}-4\, S_{2,1}, \\
\gamma^{\rm CS}_{2, 6}(N) &=& 8 \,S_5-24 \,S_{1,4}-32 \,S_{2,3}-20 \,S_{3,2}-16 \,S_{4,1}+32 \,S_{1,1,3}+24 \,S_{1,2,2}+\nonumber\\
&& + 24 \,S_{1,3,1}+20 \,S_{2,1,2}+24 \,S_{2,2,1}+8 \,S_{3,1,1}-16 \,S_{1,1,2,1}.
\ea
and this time the argument is $S_\mathbf{a} \equiv S_\mathbf{a}(N/2)$. 

These expressions do not include wrapping corrections. We shall compute the virtual scaling function from these asymptotic expressions assuming, as in ${\cal N}=4$
SYM that it is independent on wrapping.

\section{The fourth order twist-2 anomalous dimension}
\label{sec:fourth}

As remarked in~\cite{Beccaria:2009ny} and in close analogy with twist-3 fields in ${\cal N}=4$ SYM, the twist-2 anomalous dimensions involve nested harmonic sums with positive indices
only. Such a restricted maximal transcendentality Ansatz reduces a lot the complexity of the determination of higher orders. In particular, we have determined 
the 8-th loop asymptotic anomalous dimension where dressing first appear. This will be a useful ingredient for the check of our results for the virtual scaling function.
After a straightforward computation, we find
\be
\gamma^{\rm CS}_{2, 8}(N) = G^{\rm CS}_{2, 8}(N) + \zeta_3\,G^{\rm CS, dressing}_{2, 8}(N),
\ee
where
\ba
G^{\rm CS}_{2, 8} &=& 
4 (5 \,S_7-20 \,S_{1,6}-39 \,S_{2,5}-45 \,S_{3,4}-35 \,S_{4,3}-27 \,S_{5,2}-15 \,S_{6,1}+48 \,S_{1,1,5}+\nonumber\\
&& + 62 \,S_{1,2,4}+58 \,S_{1,3,3}+52 \,S_{1,4,2}+38 \,S_{1,5,1}+58 \,S_{2,1,4}+64
   \,S_{2,2,3}+\nonumber\\
&& + 68 \,S_{2,3,2}+55 \,S_{2,4,1}+44 \,S_{3,1,3}+54 \,S_{3,2,2}+52 \,S_{3,3,1}+31 \,S_{4,1,2}+33 \,S_{4,2,1}+\nonumber\\
&& + 26 \,S_{5,1,1}-48 \,S_{1,1,1,4}-52 \,S_{1,1,2,3}-60
   \,S_{1,1,3,2}-52 \,S_{1,1,4,1}-40 \,S_{1,2,1,3} + \nonumber\\
&& -58 \,S_{1,2,2,2}-58 \,S_{1,2,3,1}-32 \,S_{1,3,1,2}-42 \,S_{1,3,2,1}-32 \,S_{1,4,1,1}-32 \,S_{2,1,1,3}+\nonumber\\
&& -54 \,S_{2,1,2,2}-54
   \,S_{2,1,3,1}-38 \,S_{2,2,1,2}-52 \,S_{2,2,2,1}-42 \,S_{2,3,1,1}-18 \,S_{3,1,1,2}+\nonumber\\
&& -32 \,S_{3,1,2,1}-34 \,S_{3,2,1,1}-12 \,S_{4,1,1,1}+32 \,S_{1,1,1,2,2}+32 \,S_{1,1,1,3,1}+16
   \,S_{1,1,2,1,2}+\nonumber\\
&& + 32 \,S_{1,1,2,2,1}+24 \,S_{1,1,3,1,1}+4 \,S_{1,2,1,1,2}+24 \,S_{1,2,1,2,1}+28 \,S_{1,2,2,1,1}+\nonumber\\
&& + 20 \,S_{2,1,1,2,1}+28 \,S_{2,1,2,1,1}+12 \,S_{2,2,1,1,1}-16
   \,S_{1,1,1,2,1,1}), \\
G^{\rm CS, dressing}_{2, 8} &=& 32\, (S_{2,2} + S_{3,1} - S_{1,2,1} - 2\,S_{2,1,1}).
\ea

\section{The virtual scaling function in ABJM from weak coupling}
\label{sec:experiments}

Up to vanishing terms for $N\to \infty$, the large spin expansion of our exact anomalous dimensions can be computed by the tricks reported in Appendix~(A).
Partial results can be found in~\cite{Beccaria:2009ny}. We add here the 6 loop twist-1 and 8 loop twist-2 results. They read
\ba
\gamma^{\rm CS}_{L=1}(N, h) &=& f^{\rm CS}(h)\,\left(\log\,N + \gamma_{\rm E} +\log\,2\right) + B^{\rm CS}_{L=1}(h) + \cdots, \\
\gamma^{\rm CS}_{L=2}(N, h) &=& f^{\rm CS}(h)\,\left(\log\,N + \gamma_{\rm E} -\log\,2\right) + B^{\rm CS}_{L=2}(h) + \cdots, 
\ea
where dots denote terms vanishing for large spin. The CS scaling function is half the ${\cal N}=4$ value
\ba
f^{\rm CS}(h) &=& \frac{1}{2} f^{\nnn}(h), \\
f^{\nnn}(g) &=& 8\,g^2-\frac{8}{3}\,\pi^2\,g^4+\frac{88}{45}\,\pi^4\,g^6 -16\,\left(\frac{73}{630}\,\pi^6+4\,\zeta_3^2\right)\,g^8 + \nonumber\\
&+& 32\,\left(\frac{887}{14175}\,\pi^8 + \frac{4}{3}\,\pi^2\,\zeta_3^2 + 40\,\zeta_3\,\zeta_5\right)\,g^{10} + \cdots .
\ea
The virtual scaling function derived from harmonic sums turns out to be 
\ba
B^{\rm CS}_{L=1} &=& -12 \zeta _3 h^4+\left(\frac{8 \pi ^2 \zeta _3}{3}+80 \zeta _5\right) h^6+{\cal O}\left(h^8\right), \\
B^{\rm CS}_{L=2} &=& 4 \zeta _3 h^4-88 \zeta _5 h^6+\left(-\frac{4}{15} \pi ^4 \zeta _3+16 \pi ^2 \zeta _5+1140 \zeta _7\right) h^8+O\left(h^{10}\right)
\ea
These values must be compared with the known weak-coupling expansion valid for twist-$L$ fields in 
${\cal N}=4$~\cite{Freyhult:2007pz,Freyhult:2009my,Fioravanti:2009xt}
\ba
\label{eq:BN4}
B^{\nnn}_L(g) &=& 8\, (2\, L-7)\, \zeta_3\,g^4 + \left(-\frac{8}{3} (L-4)\, \pi^2\, \zeta_3-8\, (21\, L-62) \,\zeta _5\right)\,g^6 + \nonumber\\
&& + \left(\frac{8}{15} \,(3 \,L-13) \,\pi^4\, \zeta_3+\frac{8}{3}\,
   (11 \,L-32) \,\pi^2\, \zeta_5+40 \,(46\, L-127) \,\zeta _7\right)\,g^8+\nonumber\\
&&  + \left(-64\, (2 \,L-7)\, \zeta_3^3-\frac{128}{945}\, (11\, L-49) \,\pi ^6 \,\zeta_3-\frac{8}{45}\, (103 \,L-310)\,
   \pi^4\, \zeta _5 + \right.\nonumber\\
&& \left. -\frac{40}{3}\, (25\, L-64)\, \pi ^2 \,\zeta_7-392\, (55\, L-146) \,\zeta_9\right)\,g^{10} + \cdots.
\ea

\section{The NLO BES equation in $\nnn$}
\label{sec:N4}

Let us recall how $B_L^{\nnn}$ is computed. 
In the by now standard notation of the BES paper~\cite{Beisert:2006ez}, one has to solve the 
integral equation for the quantity $\sigma(t)$ closely related to the Fourier transform of the Bethe root density and obeying 
\ba
\label{eq:NLOBES}
\sigma(t) &=& \frac{t}{e^t-1}\,\left[K(2\,g\,t, 0)\,(\log N+\gamma_{\rm E}-(L-2)\,\log\,2)
-\frac{L}{8\,g^2\,t}\left(J_0(2\,g\,t)-1\right) + \right. \nonumber\\
&&  \frac{1}{2}\int_0^\infty dt'\,\left(\frac{2}{e^{t'}-1}-\frac{L-2}{e^{t'/2}+1}\right)\,
(K(2\,g\,t, 2\,g\,t')-K(2\,g\,t, 0)) + \nonumber\\
&&\left. -4\,g^2\,\int_0^\infty dt'\,K(2\,g\,t, 2\,g\,t')\,\sigma(t')\, \right].
\ea
The anomalous dimension is simply given by 
\be
\gamma_L^{\nnn}(g) = 16\,g^2\,\sigma(0).
\ee
The kernel appearing in the above equation is 
\be
K(t, t') = K_0(t, t') + K_1(t, t') + K_d(t, t'),
\ee
with
\ba
K_0(t, t') &=& \frac{2}{t\,t'}\,\sum_{n\ge 1}(2\,n-1)\,J_{2\,n-1}(t)\,J_{2\,n-1}(t'), \\
K_1(t, t') &=& \frac{2}{t\,t'}\,\sum_{n\ge 1}(2\,n)\,J_{2\,n}(t)\,J_{2\,n}(t'), \\
K_d(t, t') &=& 8\,g^2\,\int_0^\infty dt''\,K_1(t, 2\,g\,t'')\,\frac{t''}{e^{t''}-1}\,K_0(2\,g\,t'', t'). 
\ea
Now, the crucial point is that \refeq{NLOBES} is in close relation with the general properties of the transfer matrix 
which appear in the Baxter equation of $\nnn$ as discussed in~\cite{Freyhult:2007pz}. Since we want to derive
a suitable modification of \refeq{NLOBES} valid in ABJM, we now analyze the Baxter equation for the ABJM twist operators.

\section{Analysis of the ABJM $\mathfrak{sl}(2)$ Baxter equation}
\label{sec:baxter}

The following discussion will be done at the one-loop level will turn to be enough for our purposes.
It is the adaptation of the methods described in~\cite{KorBaxter} to the ABJM (twisted) case. The starting point is the 
Baxter $\widehat{Q}(u)$-operator where $u$ is the spectral parameter. It acts on the $L$-sites long $\mathfrak{sl}(2)$ spin chain and obeys
the (twisted) Baxter equation~\cite{Alcaraz:1988zr} 
\be
\left(u+\frac{i}{2}\right)^L\,\widehat{Q}(u+i)-\left(u-\frac{i}{2}\right)^L\,\widehat{Q}(u-i) = \widehat{t}(u)\,\widehat{Q}(u).
\ee
The operator $\widehat{t}(u)$, also known as the auxiliary transfer matrix, is a polynomial of degree $L-1$ in $u$ with coefficients
being local charges $\widehat{q}_n$ also acting on the spin chain states. The more precise form of $\widehat{t}(u)$ is 
\be
\widehat{t}(u) = i\,(2\,N+L)\,u^{L-1} + \sum_{n=2}^L\,\widehat{q}_n\,u^{L-n}.
\ee
Both the Baxter operator and the transfer matrix can be simultaneously diagonalized with the Hamiltonian. Replacing the operators with their eigenvalues,
we obtain the same difference equation for the Baxter function $Q(u)$. The suitable boundary condition is simply the requirement that $Q(u)$ is a 
polynomial of degree $N$
\be
Q(u) = {\cal N}\,\prod_{n=1}^N (u-u_n).
\ee
Its roots are readily identified with the Bethe roots since they obey the Bethe Ansatz equations. Notice that the operator $\widehat{Q}$ is required in order to 
compute the precise form of the energy eigenstates. Instead, for the eigenvalues, the polynomial $Q(u)$ is enough.

In practice, given a certain spin $N$, we can replace in the Baxter equation a polynomial Ansatz with undetermined charged $q_n$. Solving for the 
polynomial coefficients, we end with a set of algebraic equations for the charges. Each solution is associated with a specific $\mathfrak{sl}(2)$ module
appearing in the subset of $[-1/2]^{\otimes L}$ states compatible with the twisted boundary conditions. The solutions $\delta_n$ of $t(\delta_n) = 0$
are called {\em holes}. They are dual solutions of the Bethe equations whose role will be clarified in a moment.

If we look for an even $Q(-u)=Q(u)$, then the transfer matrix (eigenvalue) has also a definite parity $t(-u) = (-1)^{L-1}\,t(u)$. Thus, the 
cases $L=1, 2$ are particularly simple. For $L=1$ we do not have holes nor charges. For $L=2$ we have a single null hole $\delta=0$ and, again, no charges.
For $L>2$, non-trivial charges start to appear. As discussed in \cite{KorBaxter,Freyhult:2007pz}, the behaviour of the spin chain in the regime $N\gg L \gg 1$
is largely determined by the $N$ dependence of the holes. In particular, it is important to determine whether the holes grow large as $N\to \infty$ or vanish.
In $\nnn$, it is known that two holes have a size which is ${\cal O}(N)$ while the other $L-2$ holes are small in the sense that they vanish like $1/\log N$.

We claim that in the ABJM case, {\em all holes are ${\cal O}(1/\log N)$}. This statement can be efficiently checked by the methods of~\cite{KorBaxter}
which provide a good approximation to the holes (accurate enough for our purposes) in the $N\gg L \gg 1$ regime. The main idea
is that in the full region $u={\cal O}((2N+L)^0)$ one can solve separately the two half-Baxter equations \ba
\left(u+\frac{i}{2}\right)^L\,Q_+(u+i) &=& +t(u)\,Q_+(u), \\
\left(u-\frac{i}{2}\right)^L\,Q_-(u-i) &=& -t(u)\,Q_-(u).
\ea
where we write
\be
t(u) = i\,(2\,N+L)\,\prod_{i=n}^{L-1}(u-\delta_n).
\ee
The solution is 
\be
\label{eq:xxx}
Q_+(u) = \left[(2\,N+L)\,i\right]^{-i\,u}\,\frac{1}{\Gamma(1/2-i\,u)^L}\, \prod_{n=1}^{L-1} \Gamma(i\,\delta_n-i\,u)
\ee
with $Q_-$ being related to $Q_+$ by conjugation, regarding $u$ as real. The approximate asymptotic Baxter function is then simply
\be
\label{eq:Qas}
Q^{\rm as}(u) = Q_+(u)\,Q_-(-i/2) + Q_-(u)\,Q_+(i/2).
\ee
The request of zero residue at the holes leads to the quantization condition 
\be
\label{eq:quantization}
\delta_n\,\log(2\,N+L) + L\,\mbox{Arg}\,\Gamma\left(\frac{1}{2}-i\,\delta_n\right) + 
\sum_{j=1}^{L-1}\mbox{Arg}\,\Gamma(1+i\,\delta_n-i\,\delta_j) = \frac{\pi}{2}\,k_n,
\ee
where $k_n$ are integers. Asymptotically this means
\be
\delta_n = \frac{\pi}{2}\,\frac{k_n}{\log(N + L/2) + (2L+1)\,\log 2 + \gamma_{\rm E}} = {\cal O}(1/\log N).
\ee
We can test the accuracy of \refeq{quantization} by comparing its solutions with the exact holes $\{\delta_n^{\rm exact}\}$. The results are shown
in the Tables reported in Tab.~(1-2) for the non-trivial cases $L=3,4,5$. The agreement is very good. We thus have very strong indications that 
all the $L-1$ holes are vanishing for large spin.
As an immediate (one-loop) consequence, we can evaluate the anomalous dimension which (from \refeq{Qas}) is 
\be
\gamma_{L, 2}^{\rm CS}(N) = 4\,\log(2\,N+L)-4\,\psi(1)+2\,\sum_{n=1}^{L-1}\left[\psi\left(\frac{1}{2}+i\,\delta_n\right)
+\psi\left(\frac{1}{2}-i\,\delta_n\right)-2\,\psi(1)\right],
\ee
where the second $-4\,\psi(1)$ term comes from the different numbers of factors in the last two terms of \refeq{xxx} and is absent in $\nnn$.
Taking the large $N$ limit with $\delta_n\to 0$ we find 
\be
\gamma_{L, 2}^{\rm CS}(N) = 4\,\left[\log\,N + \gamma_{\rm E}+(3-2\,L)\,\log 2\right] + \cdots
\ee
In particular, this is in agreement with the explicit one-loop results at  $L=1, 2$.

\section{The integral equation determining $B_L(g)$}
\label{sec:newintegral}

From the results of the previous section, we can repeat the steps leading to \refeq{NLOBES}. The main differences are listed below.
\begin{enumerate}
\item In $\nnn$ SYM, the natural universal constants accompanying $\log N$ were the combination
\be
\log\,N + \gamma_{\rm E} -(L-2)\,\log\,2.
\ee
This is replaced in ABJM by the (one-loop) combination 
\be
\log\,N + \gamma_{\rm E} -(2\,L-3)\,\log\,2.
\ee
\item The explicit factor $L-2$ in \refeq{NLOBES} can be traced back to the number of small holes $N_h$ which vanish for large $N$. It becomes $L-1$
in ABJM. Similarly, there are various terms with factors $2$ which are really $L-N_h$. This is 1 in ABJM. 
\end{enumerate}
In conclusion, we can write a modified \refeq{NLOBES} depending on a parameter $\xi$ such that 
\be
\xi^{\nnn} = 1,\qquad \xi^{\rm CS} = \frac{1}{2}.
\ee
It reads 
\ba
\label{eq:NLOCS}
\sigma_\xi(t) &=& \frac{t}{e^t-1}\,\left[\xi\,K(2\,g\,t, 0)\,\left(\log \frac{N}{\xi}+\gamma_{\rm E}-\left(\frac{L}{\xi}-2\right)\,\log\,2\right)
-\frac{L}{8\,g^2\,t}\left(J_0(2\,g\,t)-1\right) + \right. \nonumber\\
&&  \frac{1}{2}\int_0^\infty dt'\,\left(\frac{2\,\xi}{e^{t'}-1}-\frac{L-2\,\xi}{e^{t'/2}+1}\right)\,
(K(2\,g\,t, 2\,g\,t')-K(2\,g\,t, 0)) + \nonumber\\
&&\left. -4\,g^2\,\int_0^\infty dt'\,K(2\,g\,t, 2\,g\,t')\,\sigma_\xi(t')\, \right].
\ea
From this equation we immediately obtain the following exact scaling relation valid at NLO in the large spin expansion
\be
\label{eq:result1}
\gamma^{\rm CS}_L(N) = \frac{1}{2}\,\gamma^{\nnn}_{2\,L}(2\,N).
\ee
As a corollary, we obtain \refeq{final}
which is indeed verified by our explicit data at $L=1, 2$.

\section{Comments}
\label{sec:comments}

In this paper, we have proved the scaling relation \refeq{result}. This is not a totally surprising result. 
Indeed, it already appeared in the discussion of a rigid circular string stretched in both $AdS_4$ and along an $S^1$ of $\mathbb{CP}^3$ 
and carrying two spins~\cite{McLoughlin:2008he}. It has also a counterpart in the case of the generalized scaling function of~\cite{Freyhult:2007pz}
as remarked in~\cite{Gromov:2008qe}.
Here, we have given a proof valid in the case of twist operators belonging to a special $\mathfrak{sl}(2)$ subsector with many similarities to 
the $\nnn$ SYM case. We have studied in details the necessary changes in the 
derivation of the integral equation for $B_L^{\rm CS}$. This required the analysis of the asymptotic properties of the {\em hole} solutions to the 
twisted Baxter equation describing this sector of ABJM. \refeq{result} is checked against the exact known anomalous dimensions of twist operators in ABJM
presented in~\cite{Beccaria:2009ny} as well as the new fourth order result for twist-2 illustrated here.
Notice also that \refeq{result1} is valid at all orders in the coupling. In particular it permits to obtain the strong coupling
expansion of the virtual scaling function from the recent results of~\cite{Freyhult:2009my} with no effort. 

As a final comment, 
let we remark that \refeq{result} answers a technical problem, {\em i.e.} the role of the phase deformation of ABJM at the level of the virtual
scaling function. This is certainly interesting, but deserves a more sound physical motivation. We mentioned in the Introduction the important 
recent developments connecting the virtual scaling function of $\nnn$ SYM to the properties of the on-shell scattering amplitudes.
This connection has not been explored in ABJM. Compared to $\nnn$ SYM, the quite different nature of the gauge theory side sets itself against the 
close integrability structure and, in our opinion, makes the problem an intriguing one.

\section*{Acknowledgments}
We thank I.~Klebanov, N.~Gromov, D.~Fioravanti, M.~Rossi, and D.~Bombardelli for comments.

\appendix
\section{Next-to-leading large spin expansion of harmonic sums}

Let us remind a simple trick which reduces the NLO expansion of harmonic sums to the evaluation of multiple $\zeta$ values.
The general large $N$ expansion of a nested sum $S_\mathbf{a}(N)$ has a singular part which is a polynomial in $\log N$, 
a constant term, and a remainder which vanishes at $N=\infty$. The degree of the singular polynomial equals the number of leading $1$
indices in $\mathbf{a} = (a_1, \dots, a_\ell)$. This singular part can be extracted by using the shuffle algebra described in~\cite{Blumlein:2003gb}.
The main point is that if we start with ($a_1\neq 1$, but it can also be absent)
\be
\label{eq:sum}
S_{1, {\scriptsize \underbrace{\tiny 1, \dots, 1}_k}, a_1, a_2, \dots, },
\ee
we can add and subtract the product
\be
\frac{1}{k+1}\,S_1\,S_{\scriptsize \underbrace{1, \dots, 1}_k, a_1, a_2, \dots, }.
\ee
Using the shuffle algebra in one of this terms, we cancel the initial sum in \refeq{sum}. Repeating iteratively this procedure
we obtain the desired result. As an example let us consider $S_{1,1,1,2}$. Applying the above algorithm, we prove that 
\ba
S_{1,1,1,2} &=& \frac{1}{6} S_2 S_1^3+\left(\frac{S_3}{2}-\frac{S_{2,1}}{2}\right) S_1^2+\left(\frac{S_4}{2}-S_{3,1}+S_{2,1,1}\right)
   S_1+\nonumber\\
&& + \frac{S_5}{6}+\frac{S_{2,3}}{3}-\frac{S_{3,2}}{6}-\frac{S_{4,1}}{2}+\frac{1}{2} S_{2,1,2}+S_{3,1,1}-S_{2,1,1,1}.
\ea
Taking the values at infinity we simply get 
\ba
S_{1,1,1,2}(N) &=& \frac{1}{36} \pi ^2 L^3-\frac{1}{2} \zeta_3 L^2+\frac{1}{40} \pi ^4 L-\zeta_5-\frac{1}{36} \pi ^2 \zeta_3 + \cdots,
\ea
where $L = \log\,N + \gamma_{\rm E}$ is the NLO expansion of $S_1(N)$.

\newpage
\TABLE{
\begin{tabular}{|l|ll|ll|}
\hline
$N$ & $\delta^{\rm ex}_{L=3}$ &   $\delta^{\rm qc}_{L=3}$ & $\delta^{\rm ex}_{L=4}$ & $\delta^{\rm qc}_{L=4}$  \\
\hline
4  & 0.22795472   & 0.22787539    & 0.40848291  & 0.40824222	   \\
8  & 0.21022561   & 0.2102162	  & 0.37840139  & 0.37832836	   \\
12  & 0.20032517  & 0.20032661	  & 0.3613073   & 0.36127663	   \\
16  & 0.1936300   & 0.19363366    & 0.3496743   & 0.34965917\\	   
20  & 0.18864721  & 0.18865113	  & 0.3409907   & 0.34098254	 \\  
24  & 0.18471752  & 0.1847212	  & 0.33413129  & 0.33412668\\   
28  & 0.18149529  & 0.18149861	  & 0.32850149  & 0.32849884   \\
32  & 0.17877816  & 0.17878113	  & 0.32375148  & 0.32374999   \\
36  & 0.17643811  & 0.17644076	  & 0.31965923  & 0.31965845   \\
40  & 0.17438935  & 0.17439172	  & 0.31607559  & 0.31607526   \\
\hline
\end{tabular}
\caption{
Comparison between the exact holes and the solutions to the quantization condition (qc) for $L=3, 4$. In the first case, 
we have two opposite holes $\pm\delta$. In the second, we have three holes $0, \pm\delta$. We show the non trivial 
value $\delta$ in all cases. The integers $k_n$ entering the quantization condition are $\pm 1$ for L=3, and $\pm 2, 0$ for $L=4$.
}
}

\TABLE{
\begin{tabular}{|l|ll|ll|}
\hline
$N$ & $\delta^{\rm ex}_{L=5}$ &   $\delta^{\rm qc}_{L=5}$ & $(\delta^{\rm ex}_{L=5})'$ & $(\delta^{\rm qc}_{L=5})'$ \\
\hline
4  & 0.15014187  & 0.15013878  & 0.57240546  & 0.57206928 \\
8  & 0.14425421  & 0.14424979  & 0.53079427  & 0.53066078\\
12  & 0.14046208  & 0.14045861  & 0.50680642  & 0.50673907\\
16  & 0.13769667  & 0.13769405  & 0.49041072  & 0.49037163\\
20  & 0.13553582  & 0.13553381  & 0.47815412  & 0.4781293\\
24  & 0.13377116  & 0.13376958  & 0.46846905  & 0.46845227\\
28  & 0.13228511  & 0.13228384  & 0.46052135  & 0.46050948\\
32  & 0.13100508  & 0.13100404  & 0.45381825  & 0.45380955\\
36  & 0.12988321  & 0.12988235  & 0.44804621  & 0.44803966\\
40  & 0.12888638  & 0.12888565  & 0.44299429 & 0.44298925\\
\hline
\end{tabular}
\caption{
Comparison between the exact holes and the solutions to the quantization condition (qc) for $L=5$. The holes are 
$\pm\delta_{L=5}, \pm (\delta_{L=5})'$. We show the positive holes. The integers $k_n$ entering the quantization condition are $\pm 3, \pm 1$.
}
}

\end{document}